\documentclass[dvips]{acta}
\usepackage{supertabular,lscape,epsfig}
\usepackage{amssymb}
\usepackage{amsmath}

\makeatletter
\def\hlinewd#1{%
  \noalign{\ifnum0=`}\fi\hrule \@height #1 \futurelet
   \reserved@a\@xhline}
\makeatother

\SetPages{0}{0}

\SetVol{0}{2016}
\begin{document}
\begin{Titlepage}
\Title{ Constraints on the formation of double neutron stars from the observed eccentricities and current limits on merger rates.}
%\Author{C~h~r~u~{\'s}~l~i~{\'n}~s~k~a, M., B~e~l~c~z~y~n~s~k~i, K., B~u~l~i~k, T., C~i~e~{\'s}~l~a~r, M. and G~{\l}~a~d~y~s~z, W.}
\Author{C~h~r~u~s~l~i~n~s~k~a, M., B~e~l~c~z~y~n~s~k~i, K., B~u~l~i~k, T. and G~l~a~d~y~s~z, W.}
{Warsaw University Observatory, Al. Ujazdowskie 4, 00-478 Warsaw, Poland}
% e-mail: m.chruslinska@student.uw.edu.pl}

\Received{Month Day, Year}
\end{Titlepage}

\Abstract{
We employ population synthesis method to model the double neutron star (DNS) population and test various possibilities on natal kick velocities gained by 
neutron stars after their formation.
We first choose natal kicks after standard core collapse SN from a Maxwellian distribution with velocity dispersion of $\sigma$=265 $\rm km \ s^{-1}$ 
as proposed by Hobbs et al. (2005) and then
modify this distribution by changing $\sigma$ towards smaller and larger kick values.
We also take into account the possibility of NS formation through electron capture supernova.
In this case we test two scenarios: zero natal kick or small natal kick, drawn from Maxwellian distribution with $\sigma = 26.5 \ \rm km \ s^{-1}$.
We calculate the present-day orbital parameters of binaries and compare the resulting eccentricities with those known for observed DNSs.
As an additional test we calculate Galactic merger rates for our model populations and confront them with observational limits.
We do not find any model unequivocally consistent with both observational constraints simultaneously.
The models with low kicks after CCSN for binaries with the second NS forming through core collapse SN are marginally consistent with the observations.
This means that either 14 observed DNSs are not representative of the intrinsic Galactic population, or that our modeling of DNS formation needs revision. 
}
{ Stars: neutron, binaries: general, supernovae: general }

\section{Introduction}

Double neutron star (DNS) systems are rare among the observed population of neutron stars.
To date we know only 14 such systems.
Before they form, they have a large chance of getting disrupted
when one of the stars undergoes a supernova explosion.
Disruption may be caused by the associated mass loss, which leads to additional velocity gained by the center of mass of a binary (called Blaauw kick (Blaauw 1961)) 
and affects its orbital parameters, or by additional velocity (so-called {\it natal kick}) gained by the newborn NS due to asymmetries developed during SN explosion (or by combination of both).
Based on observations of proper motions of young single pulsars, Hobbs et al. (2005) argue that natal kicks are of order of a few hundreds $\ \rm km \ s^{-1}$ 
and may be well described by a Maxwellian distribution with a dispersion velocity $\sigma_{0}$ = 265 $\ \rm km \ s^{-1}$.  
One of the biggest mysteries concerning DNS is how do they remain bound after two supernovae, especially if they involve substantial natal kicks.
Arguments were raised, however, that some neutron stars should have formed with small natal kicks ($\lesssim 50 \ \rm km \ s^{-1}$), particularly in Be/X-ray binaries 
(Pfahl et al. 2002).
To account for the population that apparently forms with small natal kicks, an alternative to standard core-collapse SN mechanism was proposed: the so-called electron-capture supernova (ECSN), 
where O-Ne-Mg degenerate core-collapses due to electron captures on Mg and Ne (Miyaji et al. 1980; Nomoto 1984, 1987) and leads to small %($\lesssim $50 $\ \rm km \ s^{-1}$) 
or no natal kick
(van den Heuvel 2011, Andrews et al. 2015). This mechanism is believed to operate only within a very narrow range of initial stellar masses.
This range is believed to be broadened in binary systems, 
as suggested by Podsiadlowski et al. (2004), thus it should occur almost only in binaries. \newline
In this paper we use \textsc{StarTrack} population synthesis code (Belczynski et al. 2002, Belczynski et al. 2008) to model
galactic double neutron star population as could exist today and
test various possibilities on natal kicks that were imparted on neutron stars during their formation, taking into account the possibility of NS formation through ECSN.
In each case we account for the Blaauw kick, present even in case of the entirely symmetric SN explosion.
We calculate present-day eccentricities of DNS systems and compare them to those known from available observational data.
As an additional test we calculate Galactic merger rate for each model and confront them with current observational limits.

\section{Method}
\subsection{ Model populations }
We use the \textsc{StarTrack} population synthesis code %(Belczynski et al. 2008, Belczynski et al. 2002)
to simulate model populations of double neutron stars evolving in isolation. \textsc{StarTrack} employs Monte Carlo techniques to
model the evolution of single and binary stars and to date was used many times to study the evolution of binary compact objects.
The population synthesis code allows us to calculate the parameters and age of each system at various stages of its evolution and to follow its formation history.
It also deals with SN explosions on orbits of arbitrary eccentricity and accounts for mass losses as well as for asymmetries during SNs, expressed in natal kicks that
NSs receive at the moment of their formation (Belczynski et al. 2008).
In this paper we modify the assumed natal kick velocity distribution, from which the magnitude of velocity gained by a newly formed neutron star is drawn after each supernova explosion. 
The standard distribution is described by a Maxwellian
\begin{equation}
 f(v) = \sqrt{\frac{2}{\pi}} \frac{v^{2}}{\sigma^{3}} exp(-\frac{v^{2}}{2\sigma^{2}} )
\end{equation}
 with $\sigma_{0} = 265 \ \rm km \ s^{-1}$, as proposed by Hobbs et al. (2005). Apart from this one, we test the distributions with
 $\sigma = \frac{\sigma_{0}}{2}$ leading to smaller kicks and  $\sigma = 2\sigma_{0}$, leading to larger kicks.
We also allow for NS formation through ECSN if a star forms a degenerate ONe core that eventually increases its mass to $1.38 M_{\odot}$,
when the core collapses due to electron capture on Mg and Ne. 
The detailed treatment of ECSN in \textsc{StarTrack} is described in section 2.3.1 of Belczynski et al. (2008).
ECSN are believed to lead to no or small natal kicks ($\lesssim 50 \ \rm km \ s^{-1}$), 
thus in case of this formation channel we test two scenarios: in the first one we apply zero natal kick velocity and in the second one we use
Maxwellian distribution with $\sigma = 26.5 \ \rm km \ s^{-1}$ value (with the mean velocity of $\sim \rm 40 \ km/s$). 
With smaller natal kicks, change of orbital parameters of a binary depends mostly on the mass loss during the SN. Chances for survival are then somewhat larger,
but even with zero natal kick binary may disrupt due to substantial mass loss.
After the population synthesis code execution we are left with a binary at the time when both stellar remnants have formed. Knowing their masses and orbit,
merger time of a given system can be calculated (Peters 1964).
We assume that Galactic disk was forming stars at a constant rate of 3.5 $M_{\odot} \ \rm yr^{-1}$ through last 10 Gyrs.
After the formation of a NS-NS binary, we use its properties (masses, orbital separation and its eccentricity) to calculate its orbital evolution due to
gravitational wave emission (Peters 1964). We extract the current population of Galactic disk NS-NS systems in order to compare this population with observations. 

\subsection{ Comparison with observations }

Currently there are 14 known double neutron star systems, apart from a new candidate DNS binary with e$\sim$0.6 and period of around 0.18 h (Cameron et al. in prep.).
Two of them (B2127+11 and J1807-2500) are located within globular clusters and their evolution might have been significantly different than other DNS found in the Galactic field, 
thus we are not taking them into consideration when comparing model populations with observations.
Remaining 12 binaries, together with their eccentricities, orbital periods and references providing more details on each of these systems are listed in table 1. 
In case of the recently discovered J1913+1102 only the lower limit on mass of the companion of the detected pulsar is available, which does not allow
to exclude the possibility that the companion is a massive white dwarf (Lazarus et al. 2016). 
It is included in the further analysis, noting that its exclusion (as well as inclusion of the candidate DNS) does not change our conclusions.

\MakeTable{|c|c|c|c|}{10cm}{Eccentricities and orbital periods of the observed double neutron stars}
{\hline
Name & eccentricity & orbital period [days] & reference \\ \hline \hline
J0737-3039 & 0.087 & 0.102 & Kramer et al. 2006 \\ \hline
J1756-2251 & 0.181 & 0.320 & Faulkner et. al. 2005 \\ \hline
B1534+12 & 0.273 & 0.420 & Stairs et al. 2002\\ \hline
J1829+2456 & 0.139 & 1.176 & Champion et al. 2005 \\ \hline
J1518+4904 & 0.249 & 8.634 & Janssen et al. 2008\\ \hline
J0453+1559 & 0.11 & 4.072 & Martinez et al. 2015 \\ \hline
B1913+16 & 0.617 & 0.322 & Weisberg et al. 2010 \\ \hline
J1811-1736 & 0.828 & 18.779 & Corongiou et al. 2007 \\ \hline
J1906+0746 & 0.085 & 0.166 & van Leeuwen et al. 2015 \\ \hline
J1930-1852 & 0.4 & 45.060 & Swiggum et al. 2015 \\ \hline
J1753-2240 & 0.3 & 13.637 & Keith et al. 2009 \\ \hline
J1913+1102 & 0.09 & 0.206 & Lazarus et al. 2016 \\ \hline

\hline
% \multicolumn{3}{p{8cm}}{Caption here}
 }
We test 6 different models, each characterized by the natal kick distribution assumed
for standard core-collapse SN (three possibilities:  Maxwellian distribution with $\sigma =$ 132.5, 265 or 530 $\rm km \ s^{-1}$)
and for electron-capture SN (two possibilities: 
no natal kick, or Maxwellian distribution with $\sigma = 26.5 \ \rm km \ s^{-1}$). 
For each combination of distributions $2 \cdot 10^{7}$ binaries were simulated.
We then utilize two sample Kolmogorov-Smirnov test to compare the distribution of eccentricities of observed binaries
with those obtained from simulations with different natal kick models.
For each model we focus on two subpopulations consisting of binaries formed in such a way that the second NS formed through SN of a specific type (CCSN or ECSN).
As an additional test we estimate Galactic NS-NS merger rates for each model considered in this work. In order to do so,
we follow the same approach as described in Dominik et al. (2012), performing calculations for a synthetic galaxy similar to the Milky Way
(with solar metallicity and 10 Gyr of continuous star formation at the level of 3.5 $\rm M_{\odot} \ yr^{-1}$).
We then compare our merger rates with the limits imposed by observations.

\section{Results}

Typically the formation of DNSs involves two SN\footnote{Another possible formation channel, not considered in this paper, involves accretion-induced collapse of a white dwarf to NS},
either CCSN or ECSN (with its specific natal kick, as described in section 2).
Binaries with orbits within the observed range of periods evolve through Roche-lobe overflow (RLOF) events - 
either stable mass transfer or common envelope in various combinations (see Table 4 in (Dominik et al. 2012)).
RLOF taking place before the second SN circularizes the orbit.
It implies that the distribution of eccentricities of DNS is determined by the second SN. 
Therefore our results are presented with respect to the type of the second SN. \\
The results of the simulations are summarized in Table 2. 
Clearly the dominant formation scenario was that in which the second supernova was of standard core-collapse type.
The first SN in the modeled systems is usually of electron-capture type - as it involves small or zero natal kicks,
it will not break the initially wide binary. For an ECSN to occur, mass of a star has to fall within the very specific mass
range, corresponding to the lowest-end of the mass range allowed for NS progenitors.
In the typical evolutionary path, the secondary is accreting mass during the RLOF events and hence can easily leave the range of
masses allowed for possible ECSN progenitors, favoring formation scenario requiring the second SN to be CCSN.
As before the second SN the great majority of binaries is subject to common envelope phase, which decreases the separation,
they are not so easily disrupted by the second SN as by the first one, even if it involves larger natal kicks.
When we increase natal kick magnitude (by increasing $\sigma$ value in Maxwellian distribution for CCSN natal kicks),
more binaries get disrupted during formation involving CCSN and a fraction of double neutron star 
systems formed via electron-capture SN slightly increases. 
The scenario involving formation of both NS through ECSN, for models when it ECSN to zero natal kicks, allows
for the formation of very wide binaries, with orbital periods largely exceeding the range corresponding to the observed systems. 
Such binaries would be essentially impossible to detect, we thus exclude DNS with final separations $a \gtrsim 50 AU$ from our sample -
this population almost completely vanishes when we allow for small natal kicks after ECSN.

\subsection{Distributions of eccentricities}

Table 3 shows the results of two sample Kolmogorov-Smirnov test performed to compare the distribution of eccentricities
in 12 known observed DNS (2 located in globular clusters were excluded) with distributions of eccentricities arising from our model populations of DNS.
Corresponding cumulative distributions are plotted in the figure 1.
Cases that can be ruled out at $99.7\%$ level of confidence (3-standard deviations interval), corresponding to p-values $< 0.003$, are underlined. 
Those with p-value $< 0.05$ (2-standard deviations interval) are italicized. \\
\newline
% 0-kick ECSN
In case of models with zero natal kick after ECSN, the subpopulation forming through the second SN being of electron-capture type (for all $\sigma$ values tested)
can be ruled out at 3-sigma level.
Also the subpopulation of binaries forming with the second SN being CCSN for models with $\sigma = 2\sigma_{0}$ and $\sigma=\sigma_{0}$
can be ruled out at this level of confidence, while the remaining case of binaries forming through CCSN for model with the smallest $\sigma=\frac{\sigma_{0}}{2}$ 
cannot be ruled out until 2-sigma confidence level is considered.
However, in all three cases the corresponding p-values are relatively close to the limiting value of $0.003$ separating 2-sigma and 3-sigma intervals.
\\
% small-kick ECSN
If we look at models allowing for small natal kicks after the occurrence of the ECSN,
when we require the second SN to be of electron-capture type, we find better agreement with the observed eccentricity distribution 
than in any other case that we consider (in each case p-values are close to 0.5 - exclusion of the newly discovered DNS J1913+1102 leads to even higher p-values,
in all cases above 0.8).
When it comes to the subpopulation where the second NS forms through CCSN, the models with $\sigma = 2\sigma_{0}$ and $\sigma = \sigma_{0}$ are inconsistent
with the observed distribution of eccentricities at 3-sigma level of confidence. 
Again, these two cases have p-values that are relatively close to the limiting value of $0.003$.
The remaining model with $\sigma = \frac{\sigma_{0}}{2}$
cannot be ruled out at this level of confidence (yet the corresponding p-values are $< 0.05$, thus they could be ruled out at 2-sigma).
\\
\newline
These results suggest that only the models that require the second SN in the system to
produce a small (but non-zero) natal kick, as in the case where they are chosen from a maxwellian distribution with $\rm \sigma = 26.5 \ km/s$,
can well reproduce the observed distribution of eccentricities in DNS binaries. Such small natal kicks are believed to occur in the electron-capture supernova explosion.
Introducing bigger natal kicks (even with $\rm \sigma \gtrsim 130 \ km/s$, thus smaller than standard model for natal kicks) leads to considerably worse fit to observations (see Fig.1),
however in general they cannot be ruled out at 3-sigma confidence level, unless we consider our largest natal kicks models 
($\rm \sigma = 265 \ km/s$, $\rm \sigma = 530 \ km/s$). %WRITE IT MORE FORMALLY

\subsection{Merger rates}

Basing on observations of three known Galactic DNS systems (B1913+16, B1534+12, and J0737-3039), Kim et al. (2006) 
estimated Galactic merger rate values for DNS systems to lie within the range of 3 - 190 $\rm Myr^{-1}$ .
After taking into account also the Double Pulsar (PSR J0737-3039B), Kim et al. (2015) obtained the revised range
of 7 - 49 $\rm Myr^{-1}$, with the median value of 21 $\rm Myr^{-1}$. However, % see sec. 8 in Belczynski et al. (2016), 
taking into consideration large uncertainties in the pulsar luminosity function could shift this rate up or down by an order of magnitude,
leading to merger rates ranging from 2.1 - 210 $\rm Myr^{-1}$ (Mandel \& O'Shaughnessy 2010). \\
Also recently completed first observing run of Advanced LIGO provided an upper limit on NSNS merger rate of 12 600 $\rm Gpc^{-3} yr^{-1}$ %1084 $\rm Mpc^{-3}$
(The LIGO Scientific Collaboration et al. 2016).
\\
Besides the estimates that arise directly from DNS observations, also short gamma ray bursts (GRB) may provide constraints on merger rates.
Binary compact object merger scenario is so far the most successful in explaining observational properties of short GRBs.
One should keep in mind that NS-NS mergers are only one of the possible short GRBs progenitors and currently other candidates cannot be excluded (e.g. Berger 2014). %ADD TO BIBLIOGRAPHY!! DONE
As a result, they may contribute to only a fraction of the observed events.
Based on the observed rate of short GRBs, Petrillo et al. (2013) obtained merger rates of their progenitors 
in the local universe ranging from 500 to 1500 $\rm Gpc^{-3} \ yr^{-1}$, 
which yields (assuming the local density of galaxies $\rm \rho_{gal} \eqsim 0.0116 \ Mpc^{-3}$) $\rm 43 - 130 \ Myr^{-1}$.
This result is, however, strongly dependent on the weakly constrained beaming angle of the collimated emission from the short gamma ray bursts,
as stressed by the authors (see figure 3 therein). Increasing the angle would relax the limits on merger rates. If the angle is by a factor of two
smaller than assumed by the authors, merger rate could be as high as several thousand of $\rm Gpc^{-3} \ yr^{-1}$, while increasing the angle
by a factor of two would decrease the rate to around 200 $\rm Gpc^{-3} \ yr^{-1}$.
\\
Finally, using 2 potential kilonovae observations as constraints Jin et al. (2015) estimated the local compact objects merger rate (NS-NS or BH-NS mergers)
 to be $\rm 16.3^{+16.3}_{-8.2} \ Gpc^{-3} yr^{-1}$ (dependent on the beaming angle), which translates to 0.7 - 2.8 $\rm Myr^{-1}$. 
The authors stress, however, that this estimate should be taken as a lower limit.
\\
\newline
Columns 4 and 6 of Table 2 contain our estimates of Galactic NS-NS merger rates for each of the models tested in this work,
subdivided with respect to the second SN type.
There is a clear division in terms of estimated merger rates: for each model, when the second SN was ECSN the rates are 0, while
when it was CCSN the rates are $\sim$20 - 60 $\rm Myr^{-1}$. The final DNSs from the subpopulation forming through second SN being ECSN
are too wide to merge within the Hubble time.\\
When we consider the subpopulation of DNS forming through the second SN being CCSN, these results are in general in agreement with observational constraints.
They are consistent with the revised limits obtained by
Kim et al. (2015) for all models but the one with $\rm \sigma = \frac{\sigma_0}{2}$ and zero natal kicks after ECSN. This is the highest rate 
in our results, with the value of $\rm 57.8 \ Myr^{-3}$. However, it is consistent with the broadened range for NS-NS merger rates
obtained after taking into consideration the uncertainties in the luminosity function. All of our models fall well below the upper limit
based on non-detection of signal from NS-NS merger by recently completed LIGO observing run. \\
Comparing to short GRBs based rates, only 3 of our models fall within the range estimated by Petrillo et al. (2013) 
($\rm \sigma = \sigma_{0}$ and $\rm \sigma = \frac{\sigma_0}{2}$ with zero natal kicks after ECSN and $\rm \sigma = \frac{\sigma_0}{2}$ with
small kicks after ECSN). However, if the opening angle would be by a factor of two bigger than assumed by the authors, 
the merger rates would be consistent for all of the models.
Moreover, only a fraction of DNS mergers may be responsible for GRBs and BH-NS systems are also the possible progenitors of these events,
thus GRB-based merger rates cannot be directly translated to NS-NS rates (similarly in case of kilonovae based rates).
Taking the range obtained by Jin et al. (2015) based on kilonovae observations as a lower limit, we find that all of our models agree with this estimate.
\\
\newline
In great majority of successfully formed DNS binaries the first NS formed through an electron-capture supernova, with very small natal kicks (0 - 50 km/s).
Natal kicks assumed for CCSN (from 130 km/s up to more than 800 km/s in different models) easily disrupt initially wide systems.
Thus the first supernova practically sets the final number of DNS, by what means it affects the merger rates. 
The evolutionary channel in which the second NS can form through an ECSN leads to merger rates inconsistent with
the estimates from observations of Galactic DNS, short GRBs and tentative claims of potential observations of kilonovae, 
producing binaries that are too wide to merge within the Hubble time. However, this is the product of the whole evolutionary path
and the rates are not determined just by the second SN, as in the case of eccentricity distribution.

\MakeTable{|c|c|c|c|c|c|}{11cm}{Simulation results: DNS formed with different natal kick distributions and corresponding merger rates}
{
\hline
\multicolumn{ 1}{|c|}{}& \multicolumn{1}{c|}{} & \multicolumn{ 2}{|c|}{ ECSN with zero natal kick} & \multicolumn{ 2}{|c|}{ECSN with small natal kick }  \\ \cline{ 3- 6}
\multicolumn{ 1}{|c|}{$\sigma$ } & \multicolumn{ 1}{c|}{2nd SN type} &\multicolumn{ 1}{c|}{no. of } & \multicolumn{ 1}{c|}{$\rm R_{mr}$} & \multicolumn{ 1}{c|}{no. of} & \multicolumn{ 1}{c|}{$\rm R_{mr}$} \\
\multicolumn{ 1}{|c|}{} & \multicolumn{ 1}{c|}{} &\multicolumn{ 1}{c|}{DNS } & \multicolumn{ 1}{c|}{$\rm [Myr^{-1}]$} & \multicolumn{ 1}{c|}{ DNS} & \multicolumn{ 1}{c|}{$\rm [Myr^{-1}]$} \\ \hline \hline
%66.25 & second SN: ECSN & 0.97 & 4.2e-10 & 0.21 & 0.73 \\ \hline
132.5 & ECSN & 1262 & \it{0} & 292 & \it{0} \\ \hline
265.0 & ECSN & 874 & \it{0} & 274 & \it{0} \\ \hline
530.0 & ECSN & 813 & \it{0} & 264 & \it{0} \\ \hline

132.5 & CCSN & 61853 & 57.8 & 28859 & 43.3 \\ \hline
265.0 & CCSN & 28272 & 43.5 & 13404 & 29.7\\ \hline
530.0 & CCSN & 9345 & 26.8 & 5014 & 18.7 \\ \hline

\multicolumn{6}{p{11cm}}{
The first column shows $\sigma$ value identifying distribution describing natal kick velocity gained by a newborn NS after a standard core collapse SN.
The next column specifies the type of second SN that led to formation of a DNS binary.
The next two columns correspond to models where ECSN led to zero natal kick velocity and the last two to those where ECSN led to small kick, drawn from Maxwellian distribution with
$\sigma = 26.5 \ \rm km \ s^{-1}$. 
The third and fifth columns give the final number of binaries that successfully formed a DNS and fulfilled other conditions stated in section 2 for each of the subpopulations.
Columns 4 and 5 show the Galactic merger rates $\rm R_{mr}$ calculated for models considered in this work.
Rates marked in italic indicate subpopulations inconsistent with observational constraints on compact objects mergers (see section 3.1) - these are all binaries that formed in such a way that the
second supernova in the system was of electron-capture type, regardless of the natal kick model.
}
}

\section{Discussion}

Results presented in this paper show that the subpopulation of double neutron stars which formed in such a way that the second NS in the system
was born in an electron capture SN can well reproduce the observed distribution of eccentricities in DNS systems, if small natal kick velocities are added after ECSN.
At the same time we find that this formation scenario leads to formation of binaries that are too wide (with separations a $>$ 0.1 AU,
while a system composed of typical NSs of masses 1.4 $\rm M_{\odot}$ with e $\lesssim$ 0.5 will merge within a Hubble time for a$\sim$0.01 AU)
to meet the observational limits on merger rates.
On the other hand, for the subpopulation that consists of systems where the second NS formed via CCSN we find that the merger rates are 
consistent with the observational constraints.
In this case, however, we cannot account for the distribution of eccentricities that would resemble the one produced by the observed population.
At least 99\% of binaries where the second SN was of standard core-collapse type encounter mass transfer or/and common envelope phase during their evolution after the first supernova.
Both processes lead to circularization of the orbit in our simulations.
 The distribution of eccentricities is thus determined by the second supernova 
(both natal and Blaauw kick contribute). The smaller the natal kick, the closer our result is to the observed distribution (see figure 1).
The observed distribution of eccentricities is only slightly affected by the gravitational waves emission.
Binaries stay close to the initial eccentricity for most of the time before merger and merge quickly afte they become circularized.
Thus they are observable with very small eccentricities for only a short time.
\\
\newline
This failure to meet both observational constraints simultaneously either points out a possible problem with understanding of the key
processes occurring during the evolution of double neutron stars (mass transfer, ECSN, common envelope), or indicates that the observed 
 systems are not representative of the intrinsic Galactic population.
One of the assumptions that may need further investigation is circularization of the orbit after mass transfer phases.
Observations of the eccentric mass-transferring binaries indicate that
the efficiency of tidal circularization in interacting systems may not be as high as commonly believed (e.g. Boffin et al. 2014, Walter et al. 2015). 
This problem was addressed in a recent study of Dosopoulou \& Kalogera (2016a, 2016b),
who discussed the evolution of eccentric binaries subject to different types of mass-transfer and mass loss, treated as perturbation to the general two-body problem.
It reveals that mass-transfer may lead to either decrease or increase in eccentricity over timescales
that may be shorter than the tidal timescale which acts to decrease the eccentricity.
This would add a spread to the distribution of eccentricities just before the second SN and reduce the influence of the natal kick on the final distribution,
which now in case of subpopulation where the second NS forms through CCSN determines its form. 
\\
In case of subpopulation forming with the second SN being of electron capture type we are left with binaries that are too wide to merge within the Hubble time.
This should not be because of potential progenitors of such systems do not form on narrower orbits. The initial distribution of orbital periods adopted from Sana et al. (2012) %ADD to Bibliography DONE
favors shorter periods. Great majority of binaries from this subpopulation that are located in the smaller final separations end of the present-day separation distribution (a$\sim$ 0.1 AU),
formed the first NS through ECSN and passed through the common envelope before the second SN, while the secondary was on a giant branch. 
To fit the conditions required for later formation of NS through ECSN, the binary enters CE with the very specific values of the envelope mass and secondary's core mass, 
CE is ejected with separation an order of magnitude bigger than required for the merging population 
(with relatively high values of $\lambda$ parameter describing the binding energy of an envelope - consult (Dominik et al. 2012) for details on CE treatment in the simulations). 
This process (especially the mechanism leading to ejection of the envelope), crucial for shrinking the orbital separation and forming merging binaries, is however poorly understood.
\\
Finally, there is a number of factors that may influence the mass range corresponding to the stars that can
evolve through ECSN, for which we do not account in \textsc{StarTrack}, as e.g. 
core rotation rate of the potential NS progenitor, 
C/O ratio at the end of He burning phase, the amount of convective mixing (Podsiadlowski et al. 2004; Jones et al. 2013, 2014).
Broadening the allowed mass range could increase the number of binaries forming through this type of SN and diversify their evolutionary paths, potentially affecting 
the merger rates. We will address this question in our future study.

\section{Conclusions}

We simulated 6 model populations of double neutron star binaries using the \textsc{StarTrack} population synthesis code. 
Each model was characterized by assumed natal kick distribution for electron-capture (either no natal kick or Maxwellian with $\sigma=26.5 \ \rm km \ s^{-1}$) 
and core-collapse supernovae (Maxwellian with $\sigma_{0}=265 \ \rm km \ s^{-1}$ as proposed by Hobbs et al. (2005), $\frac{\sigma_{0}}{2}$ or $2\sigma_{0}$).
We applied two sample KS test to compare eccentricity distributions arising from model populations and the one based on available observational data.
The second supernova is crucial for the final eccentricity distributions of model populations,
as almost all simulated binaries pass through common envelope or mass transfer phase before the second SN 
and we assume that it leads to circularization of their orbits - more detailed treatment of eccentricity evolution during these phases may be needed.\\
We thus distinguish two subpopulations of simulated binaries based on the second supernova type involved in the formation of the binary (CCSN or ECSN).
We find that all models involving ECSN with zero natal kicks for the subpopulation forming through ECSN, as well as both models with the largest and moderate
natal kicks after CCSN can be ruled out at 3-sigma confidence level (however the last two cases lead to p-values that are close to the edge of the 3-sigma interval).
Models with $\frac{\sigma_{0}}{2}$ are marginally consistent with observations.
Only the subpopulation that involved formation of the second NS through ECSN for all models with small natal kicks after ECSN leads
to noticeably better agreement with the observations.
These results support the recent findings of Beniamini \& Piran (2016), who analyzed the masses and orbital parameters
of known DNS to constrain the magnitudes of natal kicks gained by a newborn NS and mass losses during the second SN.
They find that in majority of the observed systems the most probable formation scenario requires small
mass loss and correspondingly low natal kick during the second SN, which suggests that the second NS formed through ECSN.
\\
\newline
We estimated Galactic merger rates for our model populations and confronted them with available observational limits from known DNS, short GRBs
and potential kilonovae observations.
For each of tested models for the CCSN subpopulation the obtained rates are  $\sim$20 - 60 $\rm Myr^{-1}$ and are consistent with the
observational constraints. At the same time DNSs from the ECSN subpopulation are too wide to merge within the Hubble time, leading to 0 merger rates, 
regardless of the model.
We fail to simultaneously reproduce both the distribution of eccentricities and merger rates consistent with observations, which indicates that 
either our modeling of their formation requires revision, or that
currently known NS-NS binaries do not provide a representative sample of the underlying Galactic population.
\\
The observed DNS sample is still very limited. Increasing the number of known double neutron stars
and determination of their orbital parameters would be very desirable to better constrain and understand their formation channels
and properties. Chances are that the future projects, as for instance Square Kilometer Array Telescope will allow to enlarge this sample.

\Acknow{\\
Authors acknowledge support from the Polish NCN grant Sonata Bis 2
(DEC-2012/07/E/ST9/01360), the Polish NCN grant OPUS (2015/19/B/ST9/01099)
and the Polish NCN grant OPUS (2015/19/B/ST9/03188).}

\MakeTable{|c|c|c|c|c|c|}{8cm}{KS test results: eccentricities from different model populations compared with observations}
{
\hline
\multicolumn{ 1}{|c|}{$\sigma$ }& \multicolumn{1}{c|}{2nd SN type} & \multicolumn{ 2}{|c|}{ ECSN with zero natal kick} & \multicolumn{ 2}{|c|}{ECSN with small natal kick }  \\ \cline{ 3- 6}
\multicolumn{ 1}{|c|}{} & \multicolumn{ 1}{c|}{} &\multicolumn{ 1}{c|}{ KS statistic} & \multicolumn{ 1}{c|}{P-value} & \multicolumn{ 1}{c|}{KS statistic} & \multicolumn{ 1}{c|}{P-value} \\ \hline \hline
132.5 & ECSN & 0.98 & \underline{\it 5e-11} & 0.25 & 0.48 \\ \hline
265.0 & ECSN & 0.98 & \underline{\it 5e-11} & 0.24 & 0.47 \\ \hline
530.0 & ECSN & 0.98 & \underline{\it 5e-11} & 0.23 & 0.52 \\ \hline

132.5 & CCSN & 0.47 & \it 0.006 & 0.45 & \it 0.01 \\ \hline
265.0 & CCSN & 0.55 & \underline{\it 0.0008} & 0.53 & \underline{\emph{ 0.0014 }}\\ \hline
530.0 & CCSN & 0.59 & \underline{\it 0.0003} &  0.56 & \underline{\it 0.0005} \\ \hline
%WITHOUT THE NEW DNS J1913+1102
%132.5 & ECSN & 0.98 & \underline{\it 3.3e-10} & 0.19 & 0.81 \\ \hline
%265.0 & ECSN & 0.98 & \underline{\it 3.3e-10} & 0.19 & 0.83 \\ \hline
%530.0 & ECSN & 0.98 & \underline{\it 3.3e-10} & 0.18 & 0.85 \\ \hline
%
%132.5 & CCSN & 0.45 & \it 0.016 & 0.43 & \it 0.026 \\ \hline
%265.0 & CCSN & 0.52 & \underline{\it 0.0028} & 0.51 & \emph{ 0.0042 }\\ \hline
%530.0 & CCSN & 0.56 & \underline{\it 0.0011} &  0.54 & \underline{\it 0.0017} \\ \hline

\multicolumn{6}{p{11cm}}{
The first column shows $\sigma$ value identifying a distribution describing natal kick velocity gained by a newborn NS after a CCSN.
Second column gives type of a second supernova (standard core collapse or electron capture SN) involved in formation of a binary.
Next two columns show KS test results for models where ECSN led to zero natal kick and the last two columns correspond to models where a star formed through ECSN gained a small natal kick, chosen from 
Maxwellian distribution with $\sigma = 26.5 \ \rm km \ s^{-1}$. KS statistic tells what is the maximum distance between two cumulative distributions of the data being compared and p-value
tells what is the probability that if the two data sets were randomly sampled from identical distributions, KS statistic would be as large as observed.

P-values smaller than $0.003$, indicating cases that can be ruled out at $99.7\%$ level of confidence, are underlined.
P-values smaller than $0.05$, indicating cases that can be ruled out at $95\%$ level of confidence are italicized.
Note that the subpopulation of binaries formed in such a way that the second SN was of ECSN type for all models assuming that formation through ECSN involved small natal kicks 
(last two columns, top three rows) lead to eccentricity distributions that are noticeably more consistent with the one arising from observations than any other that we tested. 
}
 }

\begin{figure}[htb]
\includegraphics[width=1\textwidth]{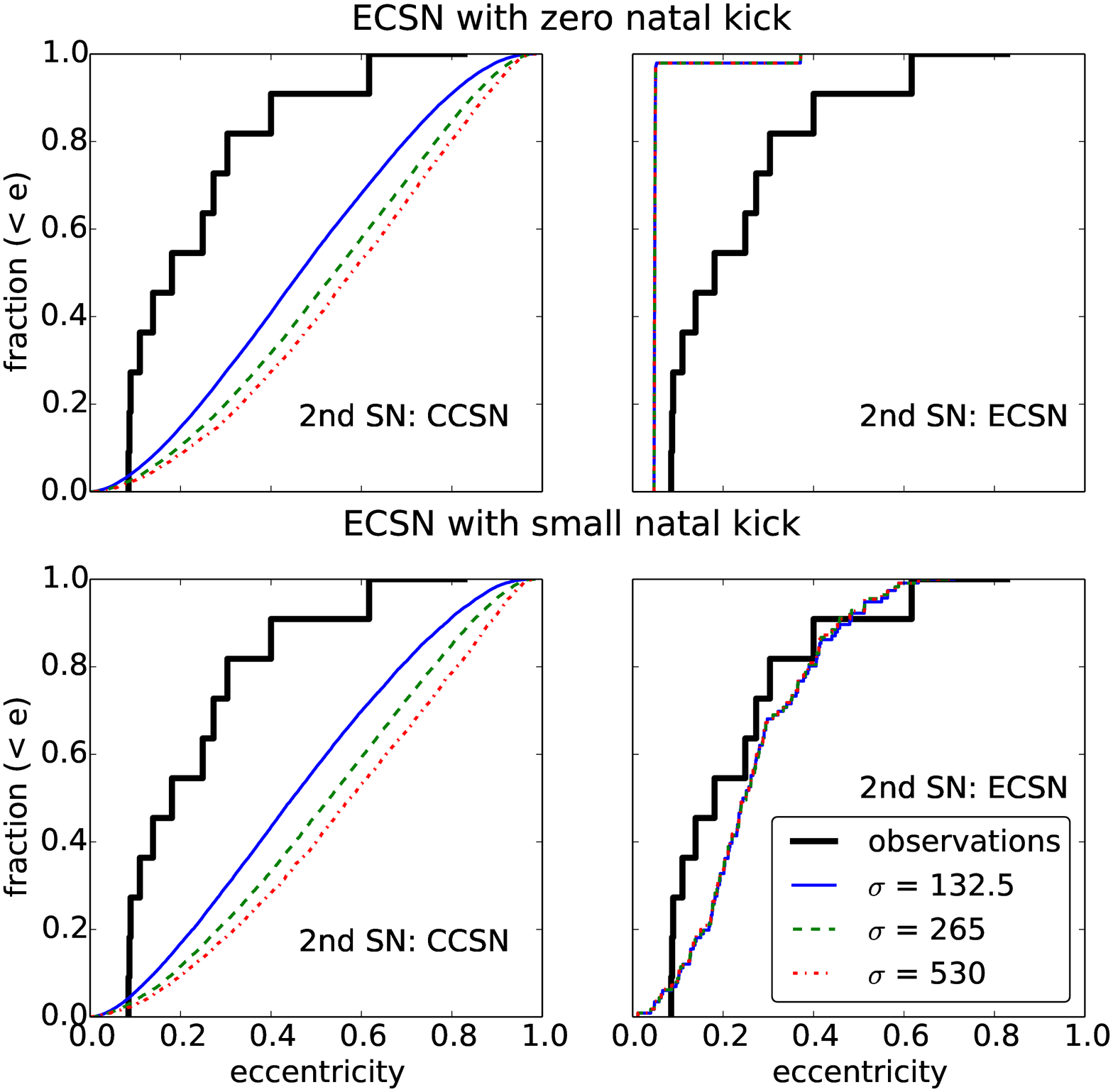}
\FigCap{

Cumulative distributions (CDF) of eccentricities in observed binaries (thick black solid line in each of the plots) and eccentricities in tested models. 
Each model is characterized by  $\sigma$ value identifying a distribution describing natal kick velocity gained by a newborn NS after a standard core collapse SN
and natal kick distribution assumed for ECSN supernovae.
Based on the second SN type (standard SN or ECSN), we distinguished two subpopulations of simulated binaries.
Left panel shows CDFs for subpopulations consisting of binaries where second SN was of core-collapse type, while right panel shows those, where the second SN was of electron-capture type.
Upper two plots represent models where ECSN led to zero natal kick, lower two plots represent models where ECSN led to small natal kick, drawn from Maxwellian distribution with $\sigma$ = 26.5 $\rm km \ s^{-1}$. 
Dashed green lines correspond to models with $\sigma_{0} = 265 \ \rm km \ s^{-1}$ for CCSN, thin blue solid lines and red dotted lines correspond to $\frac{\sigma_{0}}{2}$ and $2\sigma_{0}$ respectively.
Models with small kick ECSN for binaries formed in such a way that the second SN was of electron capture type (lower right) are most consistent with observational data, regardless of 
the natal kick distribution assumed for CCSN (see table 3 for KS test results).
}
\end{figure}

\end{document}